\documentclass[journal]{vgtc}                     %
\onlineid{0}

\vgtccategory{Research}

\vgtcpapertype{application/design study}

\title{Mixing Linters with GUIs: A Color Palette Design Probe}
\author{%
  \authororcid{Andrew McNutt}{0000-0001-8255-4258},
  Maureen C. Stone,
  and \authororcid{Jeffrey Heer}{0000-0002-6175-1655}
}

\authorfooter{
  \item
  	Andrew McNutt is with the Universities of Washington and Utah.
  	E-mail: TODO
  \item
  	Maureen Stone is with the University of Washington
  	E-mail: mcstone@cs.washington.edu
     \item 
    Jeffrey Heer  is with the University of Washington
    E-mail: jheer@uw.edu 
}

\abstract{%
Visualization linters are end-user facing evaluators that automatically identify potential chart issues. These spell-checker like systems offer a blend of interpretability and customization that is not found in other forms of automated assistance. However, existing linters do not model context and have primarily targeted users who do not need assistance, resulting in obvious---even annoying---advice. We investigate these issues within the domain of color palette design, which serves as a microcosm of visualization design concerns. We contribute a GUI-based color palette linter as a design probe that covers perception, accessibility, context, and other design criteria, and use it to explore visual explanations, integrated fixes, and user defined linting rules.
Through a formative interview study and theory-driven analysis, we find that linters can be meaningfully integrated into graphical contexts  thereby addressing many of their core issues.  We discuss implications for integrating linters into visualization tools, developing improved assertion languages, and supporting end-user tunable advice---all laying the groundwork for more effective visualization linters in any context.
}

\keywords{Linters, Color Palette Design, Design Probe, Reflection}

\teaser{
    \centering
        \includegraphics[width=\linewidth]{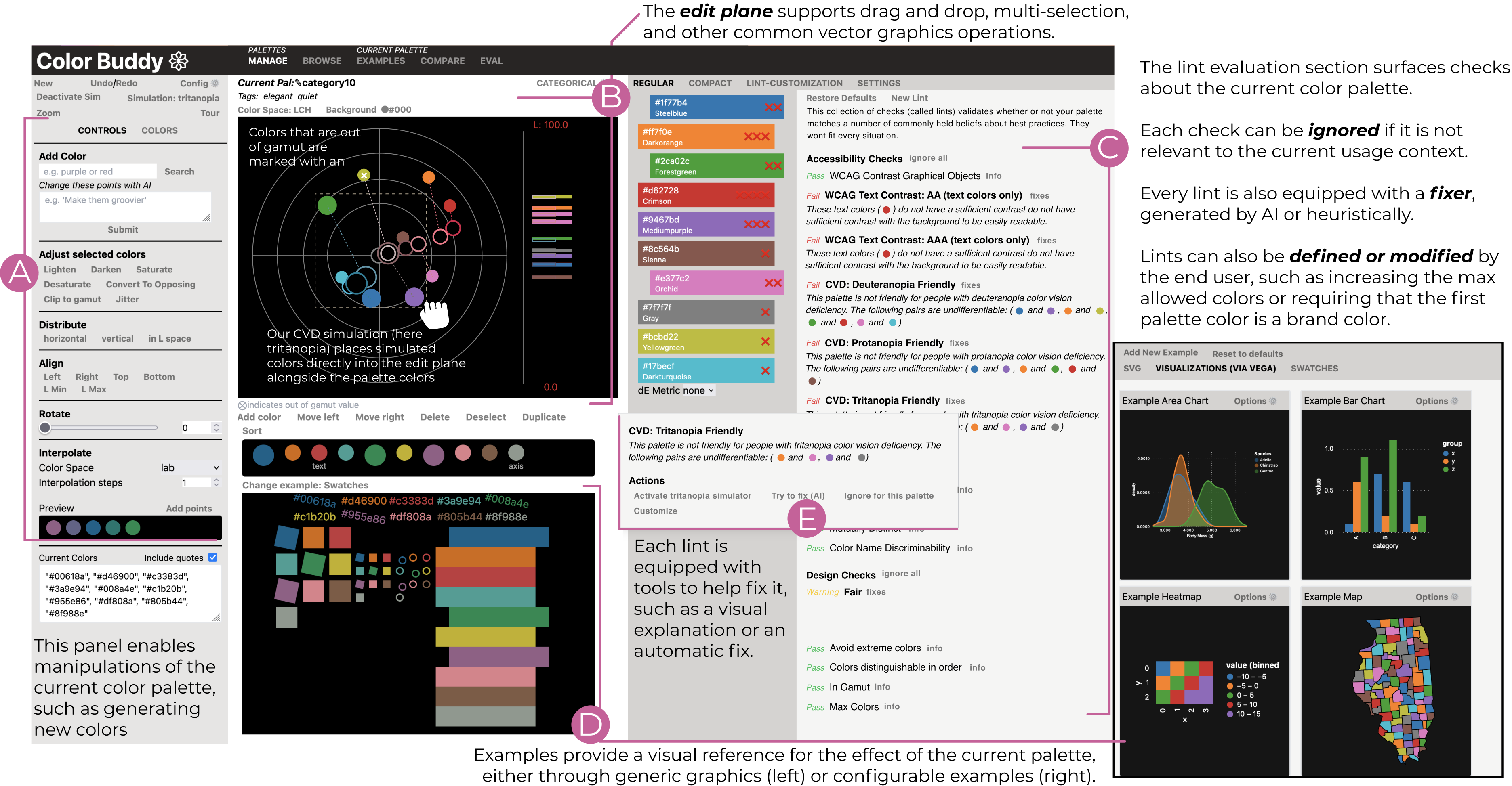}
    \caption{%
                Our design probe, \systemName{}, explores how a linter can be meaningfully integrated into a color palette design GUI.
        }
    \label{fig:annotated-guide}
}

\graphicspath{{figs/}{figures/}{pictures/}{images/}{./}} %

\usepackage{tabu}                      %
\usepackage{booktabs}                  %
\usepackage{lipsum}                    %
\usepackage{mwe}                       %

\usepackage{adjustbox}
\usepackage{array}
\usepackage{colortbl}
\PassOptionsToPackage{table}{xcolor}
\usepackage{ wasysym }

\usepackage{soul}

\usepackage[dvipsnames, svgnames]{xcolor}

\usepackage{mathptmx}                  %
\usepackage[normalem]{ulem}

\newcommand{\am}[1]{}
\newcommand{\ms}[1]{}
\newcommand{\jh}[1]{}

\newcommand{\etal}{et al.}
\newcommand{\etals}{\etal{}'s}
\newcommand{\ie}{{i.e.,}}
\newcommand{\eg}{{e.g.,}}

\newcommand{\cf}{{cf.}}
\newcommand{\ala}{\`a la}

\newcommand{\figref}[1]{\hyperref[#1]{Fig.~\ref*{#1}}}
\newcommand{\secref}[1]{\hyperref[#1]{Sec.~\ref*{#1}}}

\newcommand{\lintRule}[1]{\ul{\emph{#1}}}

\newcommand{\parahead}[1]{\paraheadd{#1}.}
\newcommand{\paraheadd}[1]{%
    \vspace{0.5em}%
    \noindent%
    \textbf{\textit{#1}}%
}

\newcommand{\inlineFig}[1]{%
    \begingroup\normalfont
    \includegraphics[height=1.2\fontcharht\font`\B]{#1}%
    \endgroup
}

\input ulem.sty\relax
\catcode`\@=11 %

\font\uwavefont=lasyb10 scaled 652
\def\uwave{%
  \bgroup
    \markoverwith{%
      \lower3.5\p@\hbox{\uwavefont\char58}%
    }%
  \ULon
}
\newcommand\colorwave[2]{{\color{#1}\uwave{\color{black} #2}}}

\newcommand{\hlc}[2][yellow]{{%
                  \colorlet{foo}{#1}%
                  \sethlcolor{foo}\hl{#2}}%
}
\definecolor{quoteColor}{HTML}{C46299}
\newcommand\qt[1]{\hlc[quoteColor!30]{``#1''}}
\newcommand{\pxx}[1]{\textbf{P$_{#1}$}}

\usepackage[normalem]{ulem}
\definecolor{linkColor}{HTML}{257E98}
\setuldepth{Berlin}
\newcommand\asLink[2]{\textcolor{linkColor}{\href{#1}{\ul{#2}}}}

\newcommand{\numLints}{35}

\newcommand{\systemName}{\textsc{Color Buddy}}
\newcommand{\languageName}{\textsc{PaletteLint}}

\newcommand{\systemURL}{\asLink{https://color-buddy.netlify.app/}{color-buddy.netlify.app}}

\newcommand{\osf}{\asLink{https://osf.io/geauf}{osf.io/geauf}} 
\begin{document}

\firstsection{Introduction}

\maketitle

\emph{How do you know what you have done is right?}
In nuanced domains (like visualization) checking new designs can be challenging.
It can require navigation of potentially conflicting design adages, easy-to-miss details, and subtle relationships between disparate components.
Reconciling these various guidelines with personal preferences can be difficult, as advice can be conflicting and desires may not be yet fully understood or quantifiable.
Various approaches~\cite{ceneda2016characterizing} in visualization try to ensure that a given program or chart is correct, or at least not bad.
These include checklists, style guides, chart metrics, theories, automated critique, and recommenders.
Each approach has trade-offs, giving agency (and responsibility) to their operator by varying degrees.

Of particular interest are visualization linters~\cite{mcnutt2018linting, hopkins2020visualint, lei2023geolinter, chen2021vizlinter, ChartLinter}.
These systems draw an analogy with static evaluation tools from software development of the same name (essentially spell checkers for code).
They are interpretable (each \emph{lint rule} checks for a single property), they are customizable (rules can be adjusted or deactivated to taste), automated (simplifying their application), are easily pliable to the grammar-based visualization systems, and can even help evaluate automatically generated code~\cite{mcnutt2023design}.
This unique blend of properties is unavailable in other forms of assistance, such as black-boxed automated recommendations.
Despite their potential, linters face several significant issues.

\parahead{Primarily Text-Based} Linters are a tool for end-users.
In visualization, they typically check the static properties of a program written in Vega-Lite~\cite{satyanarayan2016vega} or another specification language.
While useful, designing linters to target \emph{programs} rather than GUIs excludes all but those skilled in languages like Vega-Lite from this form of assistance. The number of such users is small compared to those that do analysis in tools like Excel or Tableau, leaving many unaided.
While there have been explorations of lint-style analysis in GUIs~\cite{shin2023perceptual, kristiansen2021semantic}, these checks are typically outside of the primary work cycle, excluding interaction between the linter and
the rest of the development environment.

\parahead{Impolite/annoying} Linters are often seen as annoying.
CSS Lint warns its users that it ``Will hurt your feelings* (And help you code better)''~\cite{cssLint}.
It can be a frustrating (or even impolite~\cite{whitworth2005polite}) experience seeing a huge wall of errors while you are trying to accomplish a task---particularly in formative task stages before such feedback will be useful.
While this discomfort can be useful, in that it pushes users to internalize the expressed guidelines, making it unavoidable can lead to rejecting the tool entirely~\cite{mcnutt2023study}.
Yet the basic conversational model that linting supports (in which items are iteratively dealt with, as in a dialog) is clearly beneficial and is reflected in the design of numerous tools (\eg{} AI UIs~\cite{mcnutt2023design}).
Strategies for navigating this tension are not well established, likely impeding wider usage.

\parahead{Un-useful analysis}
Finally, lints are valuable for the quality of their analysis, which has so far been limited.
Some linters~\cite{mcnutt2018linting, meeksPlotCon} include suggestions to avoid certain chart types (\eg{} the debated adage to avoid pie charts~\cite{KosaraPieCharts}),
while others~\cite{ChartLinter} merely espouse style preferences, such as avoiding redundant encodings.
Similarly, many checked errors are unlikely to appear without the user having made substantially unusual choices, which were likely intentional.
For instance, many linters~\cite{mcnutt2020surfacing, chen2021vizlinter, hopkins2020visualint} check if the y-axis is pointed in the conventional direction, to prevent issues like the infamously deceptive Florida Gun Deaths chart.
Smart defaults typically guide the designer away from this type of design, and so deciding to use it is likely a signal of designerly intent.
Finally, with one exception~\cite{lei2023geolinter}, past linters have been agnostic to analytic context (as represented by the noisy signal of chart type); instead focusing broadly on all chart forms.
\vspace{0.5em}\noindent{}This work explores these issues via a design probe in the small, yet still intricate, domain of color palette design.
Just as how color is seen as a microcosm of visual perception~\cite{palmer1999vision}, we see palette design as being representative of many of the problems in visualization.
Both domains have rich internal structures centered in graphical domains for which there are sprawling collections of guidelines and constraints to navigate.
In addition, palette design is a challenging and common task on its own~\cite{Maxene23Building}, despite a wealth of tool support (\eg{} ~\cite{jalal2015color,meier2004interactive,salvi2024colorMaker,gramazio2016colorgorical,zhou2015survey}).

To this end, we developed a color palette linter that evaluates discrete color palettes, covering categorical, diverging, and sequential types.
It can consider many usage issues, including those related to accessibility (\eg{} contrast and color vision deficiency, or CVD), stylistics (\eg{} avoiding ``ugly'' colors), and design (\eg{} ensuring that sequential palettes are perceptually sequential).
Our linter is tightly woven into its surrounding GUI called \systemName{} (also created for this study), supporting visual explanations, integrated checks, and advice that can be ignored, tuned, or summoned to task and taste.
It is powered by a domain-specific language (DSL) that uses quantifiers and predicates augmented with palette-related facilities, such as simulating CVD.
These can be created by end-users as \emph{user defined} lints, such as ensuring that a brand color is present.
See \secref{sec:supp} for code and a link to a demo.
We reflect on our probe in an interview study with design professionals (\secref{sec:formative-study}),
as well as a critical reflection~\cite{satyanarayan2019critical}   (\secref{sec:tdps}).
We contribute a characterization of the problems, strategies, and opportunities for visualization linters.
We present a typology of linter features (\secref{sec:lint-hierachy}) and identify necessary directions for additional research (Secs. \ref{sec:tdps}-\ref{sec:disco}).
By focusing on color palette design, we engage with our critiques of prior linters unburdened by a larger breadth of visualization concerns.
GUI-based systems can meaningfully weave lints into a core part of the workflow, although their integration needs to be mindful of the different parts of a task being pursued.
Having a rich body of theory to draw on supports the evaluation of complex properties---such as effective affective usage and visual accessibility.
These findings can be meaningfully returned to visualization---such as by developing theories that support complex assertions.
This work lays the groundwork for more effective visualization linters in any context.

\section{Related Work}

We draw on prior works on supporting work with colors, linting in general, and linting for visualization specifically.

\subsection{Color Tools}

There have been a wide variety of tools developed to help users pick colors, design palettes, and analyze them.
These include a wide range of tasks, covering end-user facing tools~\cite{meier2004interactive, gramazio2016colorgorical, AdobeColor, VizPalette} to automated palette generation~\cite{kita2016aesthetic, yuan2021infocolorizer, Lee13Perceptually, Hong24Cieran}.
Nardini \etals{}~\cite{nardini2021automatic} automated color map improvement is especially closely related to our notion of automated fixes, although we naively handle a generic case whereas they handle rigorously handle several specific cases.
Jalal \etal{}~\cite{jalal2015color} characterize the design space of color pickers, highlighting several axes along which tools tend to operate.
We draw on designs from many of these tools, but merely as a setting for our linting explorations.
For instance, we employ designs from both mixed-initiative~\cite{lu2020palettailor, salvi2024colorMaker, Hong24Cieran} and direct manipulation~\cite{AdobeColor, paletton, shugrina2019color, shi2023stijl} driven palette construction.
In terms of Zhou's survey of continuous color map studies~\cite{zhou2015survey}, our work falls into the ``rule-based methods'' category, \ala{} PRAVDAColor~\cite{bergman1995rule}.

More closely related to our work, are those studies that consider evaluating palettes.
Bujack \etal{}~\cite{bujack18GoodBadUgly} synthesize the space of guidelines for continuous color maps into concrete mathematical formulas.
We note that color palette design involves more nuance than merely being a simpler version of gradients, as the colors often are required to perform a variety of roles (\eg{} text or background) in addition to carrying data.
Viz Palette~\cite{VizPalette} checks if colors are discriminable for different sizes and color name discriminability---explicitly drawing on the metrics used in Colorgorical~\cite{gramazio2016colorgorical}.
Some tools provide accessibility tools, such as for CVD or contrast---although this support is limited~\cite{tigwell2017ace}.
We draw on these prior surveys to formulate our list of lints, which we check via consultation with professional designers.
Closely related, cols4all~\cite{tennekes2023cols4all} evaluates palette properties (\eg{} does a categorical palette have equally expressive colors), but is focused on palette selection rather than design.

\subsection{Linting}

In software engineering linting typically refers to tools that identify syntactic or stylistic errors in the text of a program.
While these checks are often conducted via command line tools (such as part of a test suite), they are just as often integrated into the editing environment itself, for instance highlighting parts of the code that have behaviors marked as either
as \colorwave{red}{errors} or \colorwave{Goldenrod}{warnings}.
Each assertion about the code, referred to as a \lintRule{rule}, checks for a particular behavior or property. For instance, \lintRule{max-len} enforces that all lines of code be shorter than a predefined maximum.
Beyond stylistics, lint-like systems surface type errors, check for anti-patterns (such as duplicated code not abstracted into a function~\cite{Kucherenkojscpd}), highlight inaccessible patterns, and security leaks~\cite{webhint},
among many other errors that might be easily missed.

A central usability feature of lints is that the end-user can deactivate or modify rules.
This can be on the scale of an entire project (\eg{} lines must have $\leq 120$ characters) or deactivating the rule for a particular line (\eg{} allowing a long string one time).
\emph{Allowing the user to break the rules when appropriate is essential}, as the task model espoused by a set of particular rules will not cover all use cases.

Linters are sometimes perceived as annoying until a user internalizes the code practices espoused by that linter~\cite{mcnutt2023study, lei2023geolinter}---which can be a useful way to learn a new domain.
Li \etal{}~\cite{li2023beyond} note similar properties in creativity support tools.
However, internalizing a rule is closely related to automation bias, particularly when linters are equipped with automatic fixers.
Prompting disengagement from critical thinking can be troublesome, but it can be a useful way to teach best practices.

The idea of linters has been used outside of code.
write-good~\cite{writeGood19Ford} guides its documentation authors toward the active voice,  while HaTe Detector~\cite{winchester2023hate} checks for harmful terminology (\eg{} \colorwave{red}{blacklist}).
Data Linter~\cite{hynes2017data} looks for issues in machine learning data.
ExceLint~\cite{barowy2018excelint} uses entropic program analysis to identify issues in spreadsheets.
This highlights how lints can simplify usage of complex metrics: lints are often configured by experts and employed by users.
Our work is related to these but translated to a GUI.
Textlets~\cite{han2020textlets} explores constraint enforcement mechanisms for legal writing, such as a maximum word count.
Farsight~\cite{wang2024farsight} provides a lint-like experience for identifying harms that AI prompts might cause---although they envision it through the lens of  usable security-style alerts.
These works draw from validation in end-user software engineering, which Ko \etal{}~\cite{ko2011state} survey.

\newcolumntype{R}[2]{%
    >{\adjustbox{angle=#1,lap=\width-(#2)}\bgroup}%
    l%
    <{\egroup}%
}
\newcommand*\rot{\multicolumn{1}{R{20}{1em}}}%

\definecolor{tableYesColor}{HTML}{79c996}
\definecolor{tablePartialColor}{HTML}{f8e66d}
\definecolor{tableNoColor}{HTML}{fb9478}

\newcommand{\tableYes}{\cellcolor{tableYesColor!50}Yes}
\newcommand{\tablePartial}{\cellcolor{tablePartialColor!50}Partial}
\newcommand{\tableNo}{\cellcolor{tableNoColor!50}No}
\newcommand{\colsize}{11mm}
\newcolumntype{P}[1]{>{\centering\arraybackslash}p{#1}}
\begin{figure}
    \centering
    \small
    \setlength\tabcolsep{1.5pt}
    \begin{tabular}{lc|P{\colsize}P{\colsize}P{\colsize}P{\colsize}}
        Linter               &                               & \rot{Checkable} & \rot{Customization} & \rot{Blameable} & \rot{Fixable} \\\hline
        AI Chains            & \cite{wu2022ai}               & \tableYes{}     & \tableNo{}          & \tableNo{}      & \tableYes{}   \\
        Chart Linter         & \cite{ChartLinter}            & \tableYes{}     & \tableYes{}         & \tableNo{}      & \tableYes{}   \\
        EVM Check            & \cite{kale2023evm}            & \tablePartial{} & \tableNo{}          & \tableNo{}      & \tableNo{}    \\
        GeoLinter            & \cite{lei2023geolinter}       & \tableYes{}     & \tableNo{}          & \tableNo{}      & \tableYes{}   \\
        LitVis               & \cite{wood2018design}         & \tableYes{}     & \tableYes{}         & \tableYes{}     & \tableNo{}    \\
        Mirage linter        & \cite{mcnutt2020surfacing}    & \tableYes{}     & \tableNo{}          & \tablePartial{} & \tableNo{}    \\
        Perceptual Pat       & \cite{shin2023perceptual}     & \tablePartial{} & \tableYes{}         & \tableNo{}      & \tableNo{}    \\
        VisGrader            & \cite{hull2023visgrader}      & \tableYes{}     & \tableNo{}          & \tableNo{}      & \tableNo{}    \\
        vislint\_mpl         & \cite{mcnutt2018linting}      & \tableYes{}     & \tableYes{}         & \tableNo{}      & \tableNo{}    \\
        Visualint            & \cite{hopkins2020visualint}   & \tableNo{}      & \tableNo{}          & \tableYes{}     & \tableNo{}    \\
        Visualization Linter & \cite{zheng2019visualization} & \tableYes{}     & \tableNo{}          & \tablePartial{} & \tableYes{}   \\
        VizLinter            & \cite{chen2021vizlinter}      & \tableYes{}     & \tableNo{}          & \tablePartial{} & \tableYes{}   \\
        \languageName{}      & (this work)                   & \tableYes{}     & \tableYes{}         & \tableYes{}     & \tableYes{}
    \end{tabular}
    \vspace{-1em}
    \caption{Past visualization linters provide varying levels of support for different functionalities central to the user experience of linting. }
    \label{fig:ladder-analysis}
    \vspace{-2em}
\end{figure}

\subsection{Linting for Visualization}

Visualization is an especially appealing domain for linter-style tools because its users are rarely visualization experts, there are numerous guidelines and best practices to navigate, and it is easy to make subtle mistakes~\cite{mcnutt2020surfacing}.
First described by Meeks~\cite{meeksPlotCon} in 2017, \emph{visualization linting}, applies the idea of gentle correctable assistance to visualization.
A number of linters were subsequently developed, covering matplotlib (via vislint\_mpl~\cite{mcnutt2018linting}), chart.js (via Visualization Linter~\cite{zheng2019visualization}), Vega-Lite (via Mirage linter~\cite{mcnutt2020surfacing}), and AntV (via Chart Linter~\cite{ChartLinter}).
VizLinter~\cite{chen2021vizlinter} expanded visualization linting to include both fixers as well.
These assertions generally cover stylistics or data type usage (\eg{} bar charts should use nominal data).
Qu and Hullman~\cite{qu2017keeping} provide an enriched variation of these models by developing constraints for dashboard-like systems.
We draw on the design of these systems in our work but in the domain of color palette construction.
Visualint~\cite{hopkins2020visualint} explored the visual presentation of lint failures, finding that in-situ chart annotations aided understanding of issues---a design strategy we echo.
Closest to our work is GeoLinter \cite{lei2023geolinter} for choropleth maps, which provides more specific guidance than prior linters by drawing upon a richer model of usage context. For example, GeoLinter analyses include topics such as color-aware data binning. Similarly, we narrow our domain scope to provide more useful analyses.
VisGrader~\cite{hull2023visgrader}  employs a related testing strategy to validate student submissions in a visualization course, although their tests are bound to the specific assignments in that course.

These methods have been extended beyond the specific framing of linting.
Wu \etal{}~\cite{wu2022ai} explore the use of LLM-based prompt chaining to analyze and critique Vega-Lite programs.
We also make use of LLMs, but purely as a source of generative recommendation (following Shi \etal{}~\cite{shi2023nl2color}), rather than evaluation.
Perceptual Pat~\cite{shin2023perceptual}  offers automated commentary on visualizations via computer vision-based methods. They observe that users like the validation, but struggle with some automation biases.
LitVis~\cite{wood2018design} provides lint-like guidance to the visualization design process (rather than the designed artifact) that prompts users to reflect on the values expressed in their design.
EVM Check~\cite{kale2023evm} supports end-user checking of data interpretations via expressions of statistical models, although they do not provide assertions to guide evaluation.
We build upon these prior explorations to understand the linter design space.

\section{A Ladder of Linting Support}
\label{sec:lint-hierachy}

Linters include a variety of feature types, each of which offers a different discrete level of complexity of support forming a \emph{linting ladder}.
Here, we identify common patterns of support by comparing prior visualization linters (broadly defined) with common software linters such as eslint~\cite{awesome-eslint} or SQLFluff~\cite{Cruickshank23SQLFluff}.

\begin{enumerate}
    \setlength\itemsep{-0.25em}
    \item \textbf{Checkable}: \emph{Assertions can be evaluated}
    \item \textbf{Customizable}:  \emph{Checks can be deactivated, locally or globally, and their parameters can be adjusted}
    \item \textbf{Blameable}: \emph{The cause of the issue can be identified}
    \item \textbf{Fixable}: \emph{Automated fixes can be given}
\end{enumerate}

A check from a software linter like eslint would satisfy each of these rungs. For instance, the check \lintRule{max-len} would (1) check if the property is violated---\ie{} it evaluates if all lines of code are less than a maximum allowed length.
Offer the facility to be ignored if necessary (2), such as when using a long configuration string. Describe where the error occurred, such as by highlighting the failing line in an IDE (3). And finally offer an automated fix (4), such as inserting line breaks.
These affordances allow linters to traverse the full range of Ceneda \etals{} guidance taxonomy~\cite{ceneda2016characterizing}. A lint can help orient (\eg{} this is something to watch for), direct (\eg{} this issue must be fixed), or prescribe (\eg{}  this automatic fix is right).
This underscores their potential: users can opt-in to however much help they want in a given moment.

In \figref{fig:ladder-analysis} we categorize how all known visualization linters (or closely related systems) satisfy each of these rungs of support.
For instance, early linters (\eg{} vislint\_mpl and Chart Linter) do not provide detailed feedback.
Most linters offer some degree of end-user customization, although it is often limited.
Visualint and the Mirage linter explore visual explanation (\ie{} blame) of chart issues, although the latter is unable to identify specifically why a particular aspect failed.
VizLinter and GeoLinter have fixers that can automatically address identified issues (both of which observed substantial automation bias associated with those fixers).
Contrastingly, EVM Check and Perceptual Pat are closely aligned with linters but do not actually make testable assertions---rather they are tools the user is expected to use to make mental comparisons.
LitVis has the full tooling support of VSCode, but its reflection-driven lints can not be automatically fixed.
Notably, most visualization linters skip the customization and blame rungs, forgoing essential parts of what makes linters effective end-user tools.

Additional rungs could be added to this ladder. For instance, counterfactual (or what-if) analysis  might be added after fixable.
For instance, Farsight~\cite{wang2024farsight} generates \emph{expected harms} for identified issues in AI prompts---although
this is the sole example of this type we found.
Our ladder is based on trends in observed linters (in visualization and otherwise).
We leave later ladder expansion to future work.
\begin{figure}[t]
     \centering
     \rowcolors{2}{white}{blue!10}
     \small
     \begin{tabular}{l}\hline
          Lint                                                   \\  \hline
          Affect: Dark reds and browns are not positive
          ~\cite{bartram2017affective}                           \\
          Affect: Negative palettes should not have light colors, particularly greens
          ~\cite{bartram2017affective}                           \\
          Affect: Playful affects can have light blues/beiges/grays
          ~\cite{bartram2017affective}                           \\
          Affect: Saturated not appropriate for (calm | serious | trustworthy)
          ~\cite{bartram2017affective}                           \\
          Avoid Extreme Colors
          \cite{Muth22Color}                                     \\
          Avoid Tetradic Palettes
          \cite{Muth22Color}                                     \\
          Avoid Too Much Contrast with Background
          \cite{Muth22Color}                                     \\
          Axes Should have Low Contrast with Background
          ~\cite{bartram2010whisper}                             \\
          Background Desaturation Sufficient
          \cite{Muth22Color}                                     \\
          Blue Should Nameable as Blue
          \cite{heer2012color}                                   \\
          CVD: (Deuter | Prot | Trit)anopia Friendly
          \cite{wcag3}                                           \\
          Color Name Discriminability
          \cite{heer2012color}                                   \\
          Colors Distinguishable in Order
          ~\cite{bujack18GoodBadUgly}                            \\
          Diverging Palette In Appropriate Order (Definitional)  \\
          Even Distribution in (Hue | Lightness)
          ~\cite{bujack18GoodBadUgly}                            \\
          Fair
          \cite{tennekes2023cols4all}                            \\
          In Gamut
          ~\cite{bujack18GoodBadUgly}                            \\
          Color Distinctness: (Thin | Medium | Wide) Size Objects
          ~\cite{gramazio2016colorgorical}                       \\
          Max Colors
          \cite{Muth22Color}                                     \\
          Mutually Distinct
          ~\cite{bujack18GoodBadUgly}                            \\
          No ugly colors
          ~\cite{colourlovers}                                   \\

          Prefer yellowish or bluish greens
          \cite{Muth22Color}                                     \\
          Require color complements
          (Color Theory)                                         \\
          Right in black and white
          \cite{Brown21StoneInterview}                           \\
          Sequential Palette In Appropriate Order (Definitional) \\
          WCAG Contrast Graphical Objects
          \cite{wcag3}                                           \\
          WCAG Text Contrast: (AA | AAA)
          \cite{wcag3}
     \end{tabular}
     \caption{The lints included in our design probe. Some guidelines come from multiple sources, but we prefer surveys for referential simplicity.
          Lints with common structures are merged and represented as (a | b).
     }
     \label{fig:lint-list}
     \vspace{-2em}
\end{figure}

\begin{figure*}
    \centering
    \includegraphics[width=\linewidth]{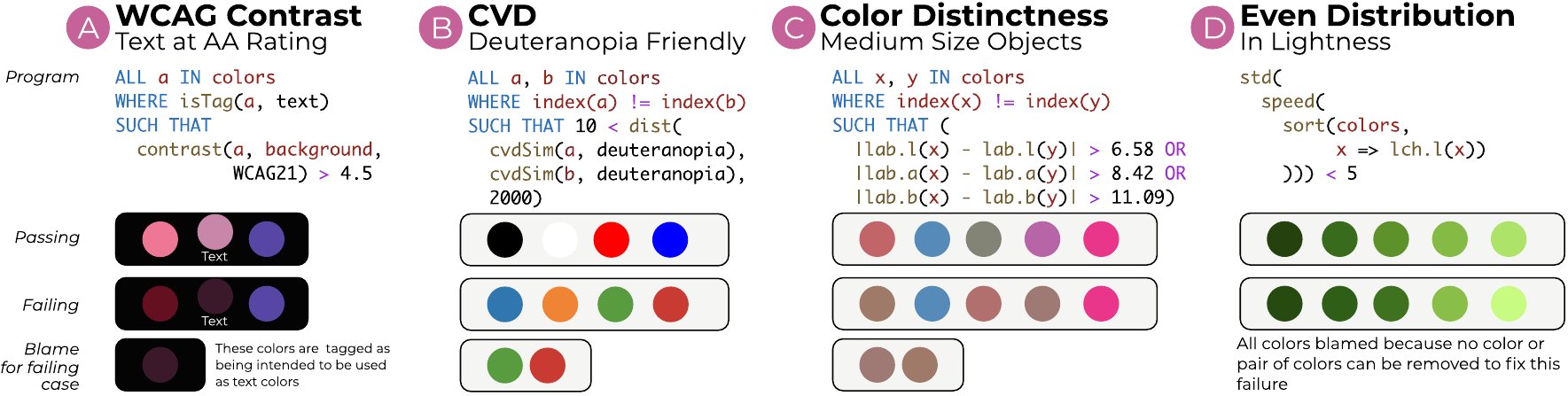}
    \caption{
        Example lints in \languageName{}'s SQL-like summary text.
        (A) checks if text objects meet contrast standards for text legibility.
        (B) compares CVD simulated color pairs using a $dE_{2000}$ heuristic~\cite{ViewSonicDeGuide} that palettes should be glanceably different.
        (C) uses values from Gramazio's d3-jnd~\cite{gramazio2016colorgorical} to test colors for distinctness as a function of size.
        (D) checks for even distribution via Bujack \etals{}~\cite{bujack18GoodBadUgly} notion of local speed (\ie{} sequential differences).
    }
    \vspace{-2em}
    \label{fig:lint-explainer}
\end{figure*}

\section{Design Probe: \languageName{}}
\label{sec:color-linting}

The central contribution of this work is an exploration of the structure of visualization linting as seen from the microcosm of color palette design.
To support this goal we conducted a design probe in which we developed a linter for color palettes called \languageName{}.
The core design loop was driven by a months-long dialogue between the first and second author (an expert in color design).
It is supported by models (\figref{fig:pal-model}) of the domain  and of the assertion space.
It is housed in a GUI called \systemName{} (\secref{sec:system}).
Its design draws on previous works, intentionally mirroring our linting ladder (\secref{sec:climbing}).

\languageName{}'s purview  is reusable discrete color palettes, as in Tableau 10 or corporate style guides.
Our task is not about picking colors, but instead about designing color palettes.
The particular values are less important than the relationships between colors and so most of the lints do not check if a specific color is included, but such analyses are possible.
This includes evaluation of properties like  accessibility (\eg{} CVD friendliness), usability (\eg{} sequential palettes being perceptually sequential), and design guidelines (\eg{} 10 colors max).

We collected \numLints{} lints (\figref{fig:lint-list}) via an informal survey of color palette design tools that provide  checks (\eg{} cols4all~\cite{tennekes2023cols4all}, Colorgorical~\cite{gramazio2016colorgorical}, and VizPalette~\cite{VizPalette}), suites of guidelines about color maps~\cite{bujack18GoodBadUgly} and palettes~\cite{Muth22Color}, as well as color perception studies~\cite{heer2012color, bartram2017affective}.
While we try to be comprehensive in this collection, we note that (like any group of lints) it will be incomplete, although a rigorous survey would be useful future work.
Beliefs about effective design will change,  accessibility measures will change~\cite{wcag3}, and new use cases will arise.
Lints address this variance by modularity: new checks can be introduced as needed.

In essence, \languageName{} is a simple interpreter that takes in a program (a lint) and data to operate over (a palette). The result of executing the program for that data is a boolean, indicating pass or fail.
Each lint also includes metadata that assists with user understanding of that result, such as a string explaining the failure, the seriousness of the error, and so on.
We include automatic fixers that draw on this information (as well as data and program) to generate fixes.
Next, we detail our model of palettes and then our model of assertions.

\newcommand{\opt}[1]{\emph{#1}}
\newcommand{\oR}{\;|\;}
\newcommand{\anD}{,\;}
\newcommand{\funC}[2]{\emph{#1}_{\small\emph{#2}}}
\newcommand{\defined}{::=}
\newcommand{\secLabel}[1]{\emph{#1}}
\newcommand{\catLabelFree}[1]{\textbf{#1}}
\newcommand{\catLabel}[1]{\hspace{-1mm}\catLabelFree{#1}\hspace{-4mm}}

\begin{figure}[t]
    \vspace{-1em}
    $$
        \begin{array}{l}
            \secLabel{Palette} = [Color] + \emph{Background Color}                                                                           \\
            \left\{
            \begin{array}{ll}
                \catLabel{Type}         & \defined{} \opt{Categorical}\oR{} \opt{Sequential}\oR{} \opt{Diverging}                     \\
                \catLabel{Context Tags} & \defined{} \{\opt{Calm} \anD{} \opt{Exciting} \anD{} \opt{Scatter plot} \anD{} \opt{...} \}
            \end{array}
            \right.\vspace{0.5em}                                                                                                            \\
            \secLabel{Color}=\emph{colorSpace}(...coordinates)                                                                               \\
            \left\{
            \begin{array}{lll}

                \catLabel{Connections}   & \defined{} & \{ \opt{tint(Color)}\anD{} \opt{tone(Color)}\anD{}                                   \\
                                         &            & \opt{complement(Color)}\anD{}  \opt{...}  \}                                         \\
                \catLabel{Semantic Tags} & \defined{} & \{\opt{Brand}\anD{} \opt{Accent}\anD{} \opt{Axis}\anD{} \opt{Blue}\anD{} \opt{...}\}
            \end{array}
            \right.\vspace{0.5em}                                                                                                            \\
            \secLabel{Lint} = \catLabelFree{Program } \emph{(\figref{fig:assertion-language})} + \emph{Required }\catLabelFree{Context Tags} \\ \emph{\hspace{0.2in}}+ \emph{Presentational Metadata} + \emph{Allowed } \catLabelFree{Types}
        \end{array}
    $$
    \vspace{-2em}

    \caption{Our model of color palettes and the lints that evaluate them.
    }
    \label{fig:pal-model}
    \vspace{-2em}
\end{figure}

\newcommand{\boolOp}{\oplus_{b}^{v, v}}
\newcommand{\mathOp}{\oplus_{n}^{n, n}}
\newcommand{\agg}{\oplus_{v}^{arr}}
\newcommand{\arrOp}{\oplus_{arr}^{e, arr}}
\newcommand{\colorOp}{\oplus_{n}^{c, c}}
\newcommand{\colorTransform}{\oplus_{v}^{c}}
\newcommand{\colorCheck}{\oplus_{b}^c}
\newcommand{\argSym}{$\theta$}
\begin{figure}[t]
    \renewcommand{\arraystretch}{1.2}
    \centering
    \vspace{-1em}
    $$
        \begin{array}{l}
            \secLabel{Expressions}                                                                      \\
            \left\{
            \begin{array}{lcl}

                \catLabel{Expressions} & e & \defined{} b \oR \lnot e \oR e \land e \oR e \lor e  \oR q  \oR{} \boolOp{}\enspace{}v\enspace{}v \\
                \catLabel{Quantifiers} & q & \defined{} (\forall \oR{} \exists)\enspace var_i \in (arr \enspace{}st\enspace{}e), \enspace{} e  \\
            \end{array}\right. \vspace{0.5em}                                             \\

            \secLabel{Values}                                                                           \\
            \left\{\begin{array}{lcl}
                \catLabel{Values}    & v   & \defined{} b\enspace{} (bool) \oR c\enspace{} (color) \oR n\enspace{} (num) \oR var \\ & & \oR \mathOp{}\enspace{}n\enspace{} n                                                       \oR{} \agg{}  arr                         \oR{} \colorOp{} c\enspace{}c \oR{} \colorTransform{} c                                            \\
                \catLabel{Arrays}    & arr & \defined{}   [v] \oR \arrOp{}\enspace{} e \enspace{} arr                            \\
                \catLabel{Variables} & var & \defined{} arr \oR{} v                                                              \\
            \end{array}\right.\vspace{0.5em}                                       \\

            \secLabel{Operations }(\oplus_{\textit{\small output types}}^{\textit{\small input types}}) \\
            \left\{\begin{array}{lcl}
                \catLabel{Val Ops}    & \boolOp{}         & \defined{} =  \oR \neq \oR < \oR >
                \oR \sim \oR ...                                                                                                                                                             \\
                \catLabel{}           & \mathOp{}         & \defined{} * \oR + \oR - \oR \%  \oR absDiff                                                            \oR ...                  \\
                \catLabel{Array Ops.} & \agg{}            & \defined{}  count \oR{} sum \oR{} min \oR{} max \oR{} std                                                                        \\
                                      &                   & \oR{} mean \oR{} first \oR{} last \oR{} extent \oR{} ...                                                                         \\

                \catLabel{}           & \arrOp{}          & \defined{} \funC{map}{\argSym} \oR{} \funC{sort}{\argSym} \oR{} \funC{filter}{\argSym} \oR{} \funC{speed}{} \oR{} ...            \\

                \catLabel{Color Ops.} & \colorOp{}        & \defined{} \funC{dist}{\argSym} \oR \funC{dE}{\argSym} \oR \funC{contrast}{\argSym}                                              \\
                \catLabel{}           & \colorTransform{} & \defined{}  \funC{toSpace}{\argSym} \oR{} \funC{cvdSim}{\argSym}       \oR{} \funC{name}{\argSym}   \oR{}  \funC{isTag}{\argSym} \\
            \end{array}\right.
        \end{array}
    $$
    \vspace{-2em}
    \caption{\languageName{}'s abstract syntax. Each lint consists of a (potentially) nested single boolean expression, $e$.
        Palette values (\eg{} background and colors) are accessed during evaluation as variables.
        \argSym{} denotes function parameters or arguments, such as contrast algorithm in \figref{fig:lint-explainer}.
    }
    \label{fig:assertion-language}
    \vspace{-2em}
\end{figure}

\parahead{Domain modeling}
At the core of our linter is a  model (per  \figref{fig:pal-model}) that shapes the space of possible expressions and usages that are possible in our probe.
This modeling is analogous to the task modeling that drives visualization tools like Vega-Lite~\cite{satyanarayan2016vega}.
Each palette is made up of a list of colors.
Both colors and palettes include contextual properties.

The palette properties are influenced by their intended usage context.
For instance, the \catLabelFree{Type} partition captures common~\cite{Maxene23Building} intended usages.
Our initial set of lints has 4 unique \emph{Categorical} lints, as well as a unique lint for \emph{Sequential} and \emph{Diverging}.
Outside of these basic usages, palettes include a huge variety of different contexts, such as being of a \emph{calm} affect or being used in a \emph{scatter plot}.
Rather than model all such usages, we instead include a notion of \catLabelFree{Context Tags}.
These sets of properties act as filters, curating what lints should be applied.
To wit, only including the rule \lintRule{Affect: Saturated Colors Not Appropriate for Calm} palettes~\cite{bartram2017affective}) to palettes marked as calm.
We demonstrate this idea through 6 tags each with a unique lint.
The background color holds a special place in the palette because it defines the visual context in which the colors are used.
Other roles might be elevated to this level of importance (\eg{} text color), but as these will not appear in every palette, we push such specification into our color model.

The colors that make up the palettes are lightweight extensions to familiar models of color (consisting of color space and coordinates) and some metadata.
Colors can serve a variety of different purposes in a palette depending on context and user.
Just as with palettes we instead capture this through a polymorphic notion of \catLabelFree{Semantic Tags}. These capture properties like ``fixed brand color'', being a type of blue, or being an axis or text color.
Among these, if there is a straightforward representation for that tag, then we include it in the edit plane scatter plot---such as ``text'' changing the text color and ``axis'' changing the axes in \figref{fig:annotated-guide}.
These properties are accessible from within the lints, supporting assertions like \lintRule{Colors Marked as Blue Should be Namable as Blue} or \lintRule{Brand Color Should Be First}.
Palettes often have internal structure, such as from color harmonies like complements or tints.
We model this as \catLabelFree{Connections} for completion but forgo implementation as it bears little impact on the analysis of our probe. Related properties can be measured via rules like \lintRule{Require Color Complements}.

\parahead{Assertion Modeling}
To support our linting we developed a formalized representation of the process of making assertions about color palettes through a lightweight domain-specific language (DSL) synonymously called \languageName{}.
It consists of quantifiers and predicates that operate over palettes and related functions, such as simulating CVD (\emph{cvdSim}),
or identifying the English name for a color (\emph{name}).
Each lint includes an assertion as a program (\figref{fig:lint-explainer}), metadata (\eg{} description, error/warning), and what parts of \figref{fig:pal-model} it applies to.
\figref{fig:assertion-language} shows the abstract syntax while the docs show the concrete JSON syntax as well as the implementation of each check (\secref{sec:supp}).

This collection of color operations was selected to support the \numLints{} lints we identified, providing some external validity of this function domain and the expressivity of this language---see language docs, (\secref{sec:supp}) for implementation.
Using this collection of operations, \languageName{} can model all checks that designers noted in our interview study (\secref{sec:formative-study}).
More operations could be added, but this set is sufficient to demonstrate
the types of problems this assertion model can address.
\languageName{} also includes functions for manipulating palettes modeled as arrays, such as aggregate ($\agg{}$) and unary array operations ($\arrOp$). In addition to familiar functions like filter or sort, $\arrOp$ includes an analog to Bujack \etals{}\cite{bujack18GoodBadUgly} local speed operator.
This supports checks like \lintRule{Even Distribution Among Lightness} via assertions like \figref{fig:lint-explainer}D.
\languageName{} can not express every possible check. For instance, Bujack \etals{}~\cite{bujack18GoodBadUgly} notion of triangle side difference for intuitive order measure can not be expressed because user defined local variables are not supported.
This can not be correctly modeled~\cite{bujack2018ordering} by naively checking for channel monotonicity as we do in our naive proof-of-concept \lintRule{Sequential Palette In Appropriate Order}.
This is a weakness of our design, but offers an approximation that can be ignored by the user when incorrect.
Similarly, it is unclear if \languageName{} can test for rainbow or cyclic color maps. However, as these maps are primarily problematic for gradients, this case is not critical to our context.
\subsection{GUI: \systemName{}}
\label{sec:system}

Linting is not a tool that is typically used alone.
In software engineering, lints are used as part of a suite of different tools that include systems like interactive development environments, unit tests, documentation, and autocomplete.
Similarly, color palette design does not occur in isolation, and so we weave our linter closely into its containing GUI, \systemName{}.
Here we detail our probe and highlight several ways in which we have integrated our linter into our GUI.

\parahead{System Details}
\systemName{} is a web-based GUI application for color palette design.
Its design draws on vector graphics tools (\eg{} Illustrator) to utilize designer's familiarity with that tool genre---analogous to how many visual analytics tools lean on data analysts' familiarity with Tableau-like systems.
Palettes are created or modified via tools on the left side of the UI.
New colors can be added by clicking on the edit plane (\figref{fig:annotated-guide}B).
Color values can be adjusted by dragging them in the edit plane or by using a WIMP-style drag-box.
Following vector manipulation tools, we include tools like align and distribute (\figref{fig:annotated-guide}A) as well as palette-specific ones like saturate or lighten.
The UI supports opportunistic reuse~\cite{brandt2009two}, analogous to how a software developer might find a piece of code on StackOverflow and then adjust it to their purposes.
To wit, a user might transform a palette from our built-in library (\inlineFig{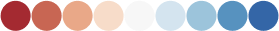}) to taste by rotating it to be  (\inlineFig{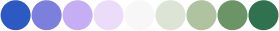}).
Next, palettes are evaluated on the UI's right side.
This includes several approaches to evaluating palettes.
There is a means to view the palette on relevant images or visualizations (\figref{fig:annotated-guide}D), to compare palettes (see appendix), and to evaluate them (our linter).
Each assessment approach comes at different stages of the task cycle.
For instance, receiving granular lint-level assessment might not be relevant while the user is still exploring the space of possible options for their task---where example images might be more useful.
This allows users to select their desired form of feedback at a given moment.
The supplement (\secref{sec:supp}) documents our probe in greater detail.

\parahead{Visual Explanations}
Next, we observe that providing a tight visual connection between the lints and the rest of the application offers opportunities to provide wider-ranging explanations.
For instance, the blames include visual renderings of the colors suggested as being at fault for a lint failure (as in \figref{fig:fix-menu}). Clicking on these circles selects them in the rest of the interface, setting them up for manipulation and correction.
We augment this by providing visual explanations of failing lints wherever possible via heuristics.
For instance, if a lint description notes a type of CVD then we provide a button to activate the appropriate CVD simulator (as in \figref{fig:annotated-guide}B).
Alternatively if a lint description characterizes something as being an error in a specific color space (\eg{} \lintRule{Even Distribution in Lightness} mentions LCH) then we include a button to switch the current palette to that space.
While these are limited, they provide straightforward ways to find support within the tool.
Future enhancement might involve automated counterfactuals or lint program visualization.

\begin{figure}[t]
    \centering
    \includegraphics[width=\linewidth]{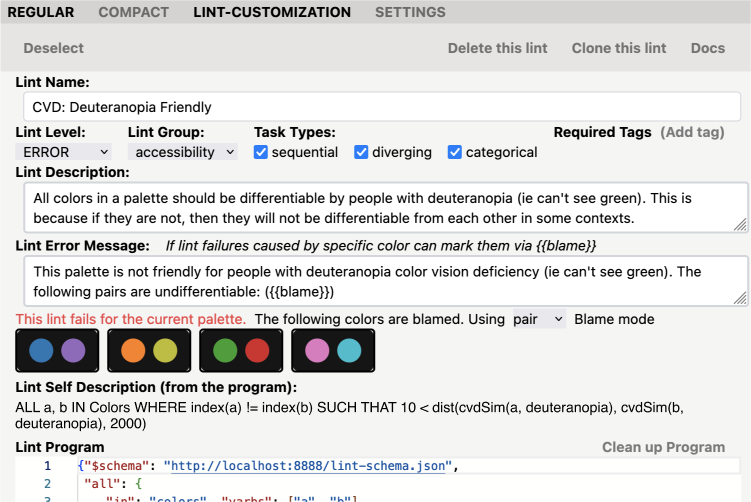}
    \caption{
        \languageName{}  lints can be configured or created by filling out a form within the \systemName{} UI.
    }
    \label{fig:user-defined-lints}
    \vspace{-2em}
\end{figure}

\parahead{User Defined Lints}
A list of pre-defined lints will rarely capture all of the constraints a user might require.
For example, a designer may wish to enforce the presence of a specific brand color but allow for some leeway for perceptually similar shades.
This flexibility is central to the philosophy of linting: rather than trying to completely model user intent, we can instead provide consistent, simple, and interpretable signals to shape one's work.
We approach this issue by developing \emph{user defined lints}---assertions defined by the end-user to achieve particular goals. We support user defined lints via end-user manipulable linting programs.
Users can manually create a new lint via a form-like UI (\figref{fig:user-defined-lints}), covering the configuration of appropriate task types, affect types, logical \languageName{} constraints, and so on.
These programs support a range of activities and intents, and allow for far more specificity than our initial list of lints (which are themselves implemented using the same mechanisms as any user defined lint).
Lints can also be customized by adjusting parameter values, such as the color contrast algorithm or discriminability thresholds.
Our probe does not support parameter adjustment outside of full program editing, but in the future, something like Ivy~\cite{mcnutt2021integrated} might simplify lint configuration.

\parahead{Visual Evaluation}
Finally, we observe that other forms of GUI-centered evaluation are possible and provide valuable support in this domain.
For instance, if a palette will only be used to create a particular chart type, then the user will be better served by evaluating how it operates on that chart type rather than all charts generically.
We support this style of analysis by enabling the user to upload example graphics either specified in SVG or Vega/Vega-Lite that are dynamically recolored---similar to Adobe Color~\cite{AdobeColor}  or Hogräfer \etals{}~\cite{revize19Hografer} modified version of ColorBrewer respectively.
The compare tab (see appendix) further supports visual evaluation via an un-editable second scatter plot that allows comparison between palettes (such as with a reference palette).
While reasonably standard~\cite{jalal2015color} these approaches dovetail with the concrete representation of correctness espoused by lints, filling in those gaps when the user is the only appropriate  oracle.
In future work, ideas like linted reflections~\cite{wood2018design} or other human-as-oracle techniques like  lineups~\cite{wickham2010graphical} could be usefully incorporated.

\subsection{Climbing the Linting Ladder}
\label{sec:climbing}

Finally, we give a tour of \languageName{} via the lint ladder.

\definecolor{quantColor}{HTML}{2F72BC}
\newcommand{\llQuant}[1]{{\color{quantColor}\texttt{#1}}}
\definecolor{varbColor}{HTML}{95261F}
\newcommand{\llVarb}[1]{{\color{varbColor}\texttt{#1}}}

\parahead{Checkable} The first rung of the ladder describes that a property can be evaluated as being true or false, which for us is captured by execution of \languageName{} programs.
A program is parsed into an abstract syntax tree following  \figref{fig:assertion-language}.
The program is executed via a traversal of the tree.
On the downwards pass variables are inserted into the evaluation environment such that operators deeper in the tree can use them.
During a loop---as in \figref{fig:lint-explainer}B's \llQuant{ALL} quantifier---the child nodes (\llQuant{WHERE} and \llQuant{SUCH THAT}) see versions of the execution environment containing the values of the variables (\ie{} \llVarb{a} and \llVarb{b}), as well as their positional indexes (\llVarb{index(a)} and \llVarb{index(b)}).
The results of these operations are bubbled upwards as return values, with each node eventually collapsing into a boolean expression, yielding a single result for the program.
Visually the results of these lints are presented to the end-user as a list, as in \figref{fig:annotated-guide}C. Each lint is shown as passing, failing, or warning.

\begin{figure}[t]
    \centering
    \includegraphics[width=\linewidth]{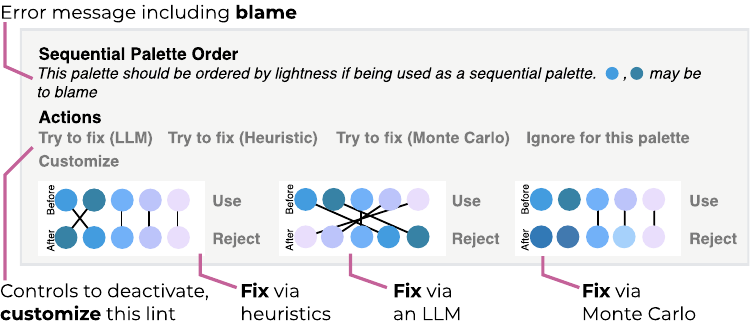}
    \vspace{-1em}
    \caption{
        Each lint comes with several forms of assistance, including checks, blames, fixes (\texttt{Try to fix}), and customizations (\texttt{Ignore}/\texttt{Customize}).
    }
    \label{fig:fix-menu}
    \vspace{-2em}
\end{figure}

\parahead{Customizable} By reifying our lints into the user interface, we make them straightforward to customize and adjust. Customization can happen either by modifying it (\figref{fig:user-defined-lints}) or by ignoring it (\figref{fig:fix-menu}).
Modifications include adjusting the description of the lint, the content of the \languageName{} program, or the conditions in which it is run (\eg{} setting it to be only run on sequential palettes).
For instance, the user could raise the  number of allowed colors on \lintRule{Max Colors}.
A critical decision to make in designing a linter is what level of granularity will be used which is connected to the depth of customization available.
For instance, VisGrader~\cite{hull2023visgrader} only evaluates specific charts in specific ways.  This is well aligned with its purpose of automatically grading d3 charts, but makes those assertions hard to re-use.
For us, the unit of input is the whole palette and the relationships contained therein, rather than individual colors, as the latter are frequently adjusted, reordered, or manipulated.
Lints can be deactivated per palette or for the entire application.
This sets our focus only on analyzing palettes rather than their usage context.
This means that we do not evaluate specific charts (as in Lee \etals{}~\cite{Lee13Perceptually} palette optimizations for choropleths) or ensure that colors are semantically resonant with a dataset (\ala{} Setlur \etal{}~\cite{setlur2015linguistic}).

\parahead{Blameable} Without a description of why a lint has failed, it can be difficult to diagnose and fix them.
Our language automatically generates a description of what colors caused a lint failure, analogous to how code lints would point to a relevant line of code.
These descriptions, or \emph{blames}, are surfaced to the end-user in the UI as in \figref{fig:fix-menu}.
We do this via a simple brute force algorithm that focuses on blaming either single colors or pairs. Respectively these support checks like  \lintRule{all colors should stand out from the background}  and  \lintRule{CVD Deuteranopia: Friendly}.
Our algorithm iterates across the color palette and checks if a new subset palette (consisting of just a single color or color pair) causes the lint to fail.
If this initial ``constructive'' approach fails, then a ``reductive'' approach is applied in which each item or pair is removed from the current palette.
Multiple colors can be blamed for a failure and so this algorithm must run to check all colors in the palette.
This algorithm could be improved by instead doing a tracing analysis in such a way that the errors can be identified analytically~\cite{acar2013core}.
While this would no doubt be faster and more robust, it is out of the scope for this work's focus on mixing linters with GUIs.

\parahead{Fixable} Finally,
there are two automated means of automatically fixing failing lints.
As in \figref{fig:fix-menu}, when presented with an error the user can select which of these three methods (labeled ``LLM'',  ``Heuristic'', and ``Monte Carlo'') they would like to use.
The LLM fixer uses external LLMs (Anthropic Haiku and GPT 4o in spring 2024) to generate fixes.
Each lint generates a failure message, which is fed into an LLM along with the current palette. The LLM proposes an update based on the prompt, which the user can accept or reject.
The heuristic fixer makes use of fixes defined outside of \languageName{} for some built-in (7) lints---\eg{} clipping to the hex color gamut for \lintRule{In Gamut} failures.
The Monte Carlo-inspired approach iteratively perturbs colors blamed for a failure until a pass is identified or a time-out is reached.
All three approaches do not always fix the error, but we find that Monte Carlo generates passing repairs more frequently than other methods, but update quality varies. See appendix for a performance comparison and more details.
Each fixer fixes a single lint error,  which can result in cycles in which fixing an error causes another error, such that auto-fixing the second error causes the first one to reappear. Tools like VizLinter~\cite{chen2021vizlinter} try to fix this through solvers.
We forgo a ``fix all'' strategy to limit the already substantial automation bias latent in presenting a fixer, although this approach may still impede creativity.

\newcommand{\pjeff}{\pxx{1}}
\newcommand{\pamy}{\pxx{2}}
\newcommand{\pbridget}{\pxx{3}}
\newcommand{\pgabby}{\pxx{4}}
\newcommand{\pjon}{\pxx{5}}

\section{Interview Study: Ecological Validity Check}
\label{sec:formative-study}

We conducted semi-structured interviews with five design professionals (denoted \pxx{x} and \qt{quoted}) who have previously made color palettes for visualization.
In particular, we sought to answer the questions: \emph{is our usage model appropriate?} and \emph{is our collection of lints aligned with perceived real-world usage?}
These interviews covered their experience making palettes and elicited their initial reactions to our probe.
Participants were identified from our social networks and were invited to participate via email. They were qualified to participate if they could be identified as having developed data color palettes in a professional capacity.
Interviews were conducted over Zoom and lasted about 60 minutes.
Participants were not compensated.

Response to our probe was generally positive, with \pjeff{} noting that \qt{we'd love to have had it} for a recent project. These responses may be biased by our connection with  participants, so we do not take their responses as evaluation. Instead, we note that their expressed usage model and expected checks are broadly in agreement with our probes.

\parahead{Usage Model}
All participants observed that designing high-quality palettes is a complex, iterative process involving numerous steps.
It requires balancing many tradeoffs to achieve a result that is functional, aesthetic, accessible, and satisfies the stakeholders.
\pamy{} summarized this view noting that \qt{the color design process is very messy.}
Participants describe palette design as consisting of a series of cycles.
Constraints are set before the work, such as requiring that a client's brand colors be used. The palette is then drafted around these constraints using a mixture of color pickers or palette generators (\eg{} Coolors~\cite{Coolors}).
Next, it is evaluated on sample graphs, relevant to the current domain. These might be found via client elicitation---\eg{} by asking  \qt{are they never making maps?} (\pjon{})---or by using \qt{a pretty large test suite} (\pjeff{}) for an in-house tool.
Then another (typically external) set of tools is used to check for accessibility---\eg{} contrast and CVD.
Finally, it is given to stakeholders for feedback.
Each of these steps are iterated many times to incorporate comments and balance conflicting requirements.  \pjeff{} observed that a good palette can take months, and \pbridget{} noted that \qt{it takes a lot more time than it should.}

This model is closely related to the iterative cyclic checks and evaluation on examples housed in our probe.
\pjeff{}, \pbridget{}, \pamy{}, and \pjon{} valued consolidating the many components of palette design in a single tool, if for no other reason than that it minimizes tool switching.
\pbridget{} noted that \qt{the more tools I have to jump to the more frustrated I get}.
This consolidation offers synergistic integration between elements, for instance allowing visual explanations of lint failures (\eg{} our CVD simulation).

However, our approach did not fit all usage models.
\pgabby{} noted that she \qt{can't usually play around with the colors} as they are usually predefined by the client. This meant that assembling a categorical palette was similar to evaluating various combinations of colors from predefined sequential palettes.
Based on our interviews and other tools (\eg{} Coolors~\cite{Coolors}, Viz Palette~\cite{VizPalette}) this model seems to be less common, however our probe could be modified to allow this more discrete form of design.
Participants noted that palettes are often components of color systems, and so can not be evaluated by themselves as our system assumes.
To wit, in visualization this includes axes and gridlines (\pbridget{}), text labels (\pjon{}), other palettes (\pjeff{}), and so on.
We address some of these via Semantic Tags, which allow us to evaluate components independently (as in \lintRule{Axes Should have Low Contrast with Background}), but managing large collections of lints is non-trivial.

\parahead{Expected Evaluations}
Participants were generally enthusiastic about having lints specifically for color palettes.
To wit, \pjeff{}, \pgabby{}, and \pjon{} had developed or modified ad hoc tools to simplify parts of the checking process---highlighting that our lints may be able to fulfill an unmet need.
\pjeff{} and \pjon{} stressed that our lints serve as an educational resource, with  \pjon{} noting that a lay user \qt{could learn all the key things they need to know, but they don't know}.
Our lint's reporting style is not unfamiliar, with all participants comparing it to prior tools, such as Viz Palette's~\cite{VizPalette} Color Report and contrast tools~\cite{webAIM, apca}.
\pjon{} suggested that our probe surpasses those works by offering automated fixing.

Participants expressed several different properties as being important to check.
All participants expressed the need to pay attention to and include brand colors.
Similarly, all participants highlighted the need to consider CVD in their designs.
\pamy{} and \pgabby{} emphasized the importance of contrast compared to CVD, as the latter is rarer but more studied due to its prevalence in white men.
\pjeff{}, \pbridget{}, and \pgabby{}  had considered ensuring that colors looked good on different devices but did not have a formal means of evaluating, other than trying out those devices.
We capture a related quantity to this via \lintRule{Out of Gamut}. \pgabby{} and \pbridget{} sometimes check grayscale to circumvent this issue, which we check via \lintRule{Right in Black and White}.
\pjeff{} noted that he tried to ensure that colors had straightforward name discriminability, which we assess via \lintRule{Color Name Discriminability}.
When asked to ideate on potential checks, participants augmented this list with checks for other visual disabilities such as \qt{astigmatism} (\pbridget{}), client preferences about particular colors (\pjon{}), color harmonies (\pbridget{}), and interactions with specific fonts (\pjon{}).
\pbridget{} went on to offer several guidelines, such as \qt{try not to do white backgrounds on mobile, because it's just glaring} and \qt{stay away from pure black}.
\languageName{} supports many of these suggestions, such as via tags like ``mobile'' and creating purpose-specific lints. Others are out of gamut (such as astigmatism or font interaction) because there is not a sufficient model to draw on to test those properties.

Our reification~\cite{beaudouin2000reification} of lints as end-user configurable entities seems to be well aligned with  participant expectations.
For instance, \pbridget{} highlighted the need to be able to adjust the values in a rule like \lintRule{No Ugly Colors} to address situational changes, such as audience or background color.
Even straightforward lints like contrast checkers might necessitate adjustments: while \pamy{} stressed the value of WCAG (referring to the contrast model used by the Web Content Accessibility Guidelines~\cite{wcag3}), \pgabby{} preferred APCA~\cite{apca}. Rather than taking an opinionated stand about which algorithm is most effective, our probe supports all common ones.
Similarly, the natural modularity of linting seemed to match expectations about palette design, with \pgabby{} observing that  \qt{it's definitely an evolution, and I'm sure it's going to change a lot. We'll have a different practice in like five years, even hopefully, it changes.}

Of course, there are limits to lints in this context.
For instance, the guidelines they implement are only effective if those guidelines are expressible. To wit, \pbridget{} noted that she had \qt{a very hard time articulating this as the general rule} for what makes a palette effective.
Using something like the espoused-preference model explored in CIERAN~\cite{Hong24Cieran}  may be valuable, although this may forgo explainability.
\pjon{} observed that color usage is situational and impedes general-purpose analysis, musing \qt{I'm gonna put text inside the bubbles in my scatter plot, do I need to use white text or black text?...that doesn't seem like a simple check.}
Our model of evaluating palettes, rather than the in-situ effect of palettes, precludes this type of contextual analysis. Future work might explore a richer model of color interaction, but our simple model seems appropriate to probe visualization linters.

\section{Analysis: Technical Dimensions of Vis Linters}
\label{sec:tdps}

Next, we use our probe to reflect on visualization linters via a critical reflection~\cite{satyanarayan2019critical}.
We structure this discussion via the Technical Dimensions of Programming Systems~\cite{jakubovic2023technical} (TDoPS), which is a descriptive framework that considers how programs are used and created across clusters of dimensions (\textbf{\emph{Denoted}} and \emph{characterized}).
We use TDoPS because its focus on the UI around programs is aligned with the artifact-as-program perspective elicited via the linting metaphor.

\parahead{Interaction \& Feedback Loops} \emph{Which loops in the system are overlapping and how far apart are the corresponding gulfs of evaluation?}
An essential missing piece from many prior visualization linters is a model of the relationship between the linter and the rest of the workflow: when is it used, how does it connect to its surroundings, and so on.
Our probe explored an integration that emphasizes connection to the surrounding context in a way that supports palette design and, per our interview, is closely aligned with the natural palette design cycle our participants use.
For example, connecting lints to both explainable notions of blame and subsequent fixes can make it easier to address a lint violation.
Considering the context of lint usage is thus valuable and can guide the types of checks to include.
In contrast, prior visualization linters have had task models like ``editing a Vega-Lite program''. While this is a common task, it is rarely an end unto itself.
It may be more effective to consider lints that arise in larger tasks, such as data exploration (\eg{} data ambiguity errors) or presentation (\eg{} stylistic judgments).

Within such contexts, the feedback loop tightness is generally prescribed by the medium.
Code linters are understood as offering nearly instantaneous feedback, although the timing of that feedback has been observed as being an important component of the perceived annoyingness of the design~\cite{mcnutt2023study}: too fast and they will be perceived as abrasive and over-eager, too slow and they will be seen as out of touch.
Our probe centers on nearly real-time feedback, as whenever a change is made the lints quickly update.
This allows for tight cycles of viewing and adjustment, which can be useful for internalizing the rules~\cite{mcnutt2023study}.
In contrast, it may be useful to explore slower feedback for some topics, \ala{} Perceptual Pat~\cite{shin2023perceptual}.
Slow checks occur naturally as a part of any design process (for color or visualization), such as by talking to stakeholders (as noted by all our participants).
Characterizing the design space of such checks is valuable future work for expanding linter utility.

\parahead{Errors} \emph{What are they and how are they handled?}
Surfacing errors in the domain of analysis is the  core of the linting interaction loop.

The quality of our lints is supported by the quite substantial body of knowledge from color perception and color design practice.
For instance, being able to check for an \lintRule{Even Distribution in Lightness} (\figref{fig:lint-explainer}) draws from the well-foundedness of the luminance and Bujack \etals{}~\cite{bujack18GoodBadUgly} formulation of local speed.
Like many before us~\cite{chen2017pathways}, we highlight that there is a need for greater theoretical development to make comparable assertions in visualization.
Developing an evaluative theory operable by humans (for interpretability) and computers (for automation) is important for making linters that can analyze more complex properties in more contexts (\eg{} overplotting or readability).
Such theories are not necessary for assembling sets of effective lints, but may be useful for identifying new ones and managing those collections.

Like prior work, we naively surface lints to the user as a long list of errors as in \figref{fig:annotated-guide}C.
This can be potentially overwhelming,  which \pjeff{} summarized by noting \qt{there's a lot going on here}.
In contrast, code linters do not present a wall of annoying errors within an IDE, in part because the edit plane of the IDE is so large.
A color palette (or a visualization) is a relatively small visual object compared to a code base, and so the errors can crowd the presentation.
Prior in-situ solutions~\cite{hopkins2020visualint} likely do not scale beyond a couple of errors. For instance, stacking multiple CVD simulations (\figref{fig:annotated-guide}) could quickly lead to clutter.
Developing means of ranking and hierarchically presenting errors is an important area for future work.

Beyond linting's natural automation biases (particularly those with fixers~\cite{lei2023geolinter}),
these checklist-like presentations can give rise to perceived false negatives: if no issues appear then the user may falsely equate this with a conclusion that there are no issues with the palette.
We contend that it is critical to calibrate user's trust in the lints so that they feel empowered to disregard lint advice. This is similar to how scaffolds in creativity support tools help users bootstrap knowledge until they are ready to creatively break the rules~\cite{li2023beyond}.
Our probe attempts this with text labels that highlight that lints are imperfect (\figref{fig:annotated-guide}C top), but the efficacy of these strategies remains unassessed.

\parahead{Notation} \emph{What notations are present and how do they interrelate?
}
In this and other linters, there are two primary notations: that of the thing being analyzed (visualization, code, color palettes) and that of the mechanism of analysis and assertion (typically heuristic checks, AST traversals, and our \languageName{} respectively).
Between these two families, assertions are typically not given as much attention.
Our exploration of the form of that assertion structure is a component of our contribution---in particular showing that a bespoke domain-centered model of assertions is possible and sufficiently expressive to capture many rules.
To that end, while we found using the language of quantifiers and predicates to be effective, metaphors like search (\ala{} SQL) or if-thens (\ala{} RulePad~\cite{mehrpour2020rulepad}) might have been successful.
The most closely related prior approach is the use of Answer Set Programming (ASP) in VizLinter and the recommender Draco.
Prior work~\cite{yang2023draco} has noted that ASP can be difficult to develop and understand, sometimes necessitating secondary tools for interpretability~\cite{schmidt2023visual}.
\languageName{}'s assertion style operates differently than ASP---in that modeling and assertion are fully distinct, rather than being interwoven through ASP modeling---but it is not yet clear if this proves more ergonomic than ASP.
We highlight that this reification~\cite{beaudouin2021generative, beaudouin2000reification} of an assertion language opens interesting new doors to wrapping user interfaces around these programs in easy-to-understand ways. We highlight usability evaluation  of these languages as important future work.

\parahead{Conceptual Structure} \emph{What is the shape of the notations at play and how do they relate to user goals?} The structure of a linter is a simple one: something is checked as being right or wrong. We found that this model works well with color palettes specifically and graphical interfaces in general, as it can be expressed through a visual presentation that is straightforward to interpret.
Moreover, we found that, for designers, this interaction was familiar as all our participants had previously used tools that contained checks of some kind---although not to the extent that \languageName{} offers.
While powerful, this evaluation-based perspective can be limiting.
For instance, it does not include support for non-evaluative categorization, as all observed behaviors are either marked as passing or failing. In contrast, cols4all~\cite{tennekes2023cols4all} evaluates if a palette is ``vivid.'' While this could be modeled as an ignorable lint (\eg{} \lintRule{Palette Should Be Vivid}), requiring opt-out imposes different cognitive burdens than categorizing labels.
Potentially useful for visualization linting, are past modifications to lint structure like \emph{neutral} or \emph{info} tests~\cite{githubCI} or embedding tutorials into linters~\cite{totalTypescript}.

Linting by itself does not support generative interactions, such as to overcome the blank page problem~\cite{satyanarayan2019critical}.
We circumvent this issue by designing for opportunistic re-use~\cite{brandt2009two} (\eg{} start from a well-formed palette)---which is aligned with palette design practice: \qt{for me it's a lot of borrowing and cheating}(\pjon{}).
Involving similar design patterns seems to be a productive means to capture needs not met directly by linters (such as those addressed by generative solutions).
An unexplored use of lints would be to use them as selectable components of a generative system---highlighting the reciprocal nature between evaluative and generative tools  (\eg{} recommenders~\cite{Hong24Cieran, salvi2024colorMaker}).
The essential differences between these tool types are not well understood in terms of results or user experience (\eg{} is one more annoying?).
This work does not interrogate these questions, but in future work we intend to use \systemName{} as a platform to explore them.

\definecolor{babyPuke}{HTML}{C3B03B}

\parahead{Customization} \emph{How can programs be modified?} A key experiment in our system is the reification~\cite{beaudouin2000reification} of lints as end-user manipulable programs; that is, we make them deeply customizable. Beaudouin-Lafon \etal{}~\cite{beaudouin2021generative} argue that such a formalization of a concept is useful as a means to empower the user to understand and create with a tool.
Subjectively, we find that this integration makes it straightforward to experiment with new types of assertions, as the palettes can be adjusted and results checked.
More concretely, our probe shows that this degree of customization is possible in a GUI-based linter and thus suggests that similar strategies might apply to visualization linters more generally.
This seems to be especially appropriate for our domain. For instance \pbridget{} observed that a particular color {\color{babyPuke}$\newmoon{}$}, \qt{works really well on dark, but you cannot use it on white because it really does look like baby puke}, highlighting the need to be able to adjust rules like \lintRule{No Ugly Colors} to context.
Lints do not need to be modeled as programs, but this form does seem to be a natural way to offer extensive customization.
Software linters can be tuned by adjusting properties like max line length. Similarly, our lints might be combined to express multiple functions through configuration, for instance a \lintRule{Sufficient Contrast with Background} rule capturing each possible algorithm and threshold.
However, this freedom may make those rules difficult to configure or discover. Instead,  relevant configurations might be materialized as separate lints (\eg{} \lintRule{APCA Contrast} vs. \lintRule{WCAG Contrast}).
Balancing lint customization and convenience is an area for future work.

\parahead{Complexity} \emph{How is complexity dealt with through design and automation?}
Complexity in linters comes from their use and management.
Our linter hides substantial complexity behind its seemingly simple pass/fail checks in the GUI.
Substantial domain knowledge may be surfaced as direct assertions (\eg{} \lintRule{CVD: Deuteranopia Friendly}).
However, as \pjon{} noted about Viz Palette~\cite{VizPalette}, fixing the identified issues can be difficult without relevant experience.
Our probe approaches this via visual explanations (such as our CVD simulation, \figref{fig:annotated-guide}B) and automated fixers.
Moreover, integration with a GUI makes it possible to not only report but also help fix the problem identified by lint.
Our linter, as with any, is driven by a growing body of knowledge.
Managing large collections of knowledge is difficult, and arises in systems like Draco~\cite{yang2023draco, schmidt2023visual} as well.
This modularity can be positive. Per \pamy{} and \pgabby{} what makes an effective palette will change.
Linters accommodate these changes by adding new lints or replacing old ones as needed.
However, growing a vast body of lints may be unsustainable, as rules may be conflicting, have differing assumptions, or may require protracted maintenance to keep up-to-date.
This is analogous to the problems that gave way to the second AI winter when the limits of expert systems became manifest~\cite{kautz2022third}.
We again suggest that substantial theoretical innovation may address some of these issues.
Being able to derive (or at least approximate) knowledge bases could make their maintenance and development simpler.

\parahead{Adaptability} \emph{What socio-technical (\eg{} learnability) dimensions are considered?}
Linters are social objects.
For instance, the documentation for the SQL linter SQLFluff~\cite{Cruickshank23SQLFluff}  includes guidance on navigating the integration of this type of validation into teams.
Enforcing a set of styles ossifies a set of beliefs about how things should work, and should be done carefully to align with tasks and teams.
In our linter we allow the user to add or remove lints as they wish, but team-driven collections of lints (as in eslint~\cite{awesome-eslint}) could support different situations or standards.
Similarly, no technical object operates in complete isolation; rather, it is supported by the surrounding ecosystem of tools, such as auto-complete and documentation.
However, these elements do not come for free.
For instance, we designed \languageName{} as a JSON DSL because extant tooling supports in-situ documentation, auto-complete, and type hints at the cost of a more cumbersome syntax.
Similarly, we surface the implementation of all the built-in lints both for documentation and for opportunistic reuse~\cite{brandt2009two}.
While these features are likely helpful for novices~\cite{mcnutt2023study}, they do not necessarily make our DSL easy to learn which likely has a substantial learning curve associated with it as it captures diverse concepts like predicate logic and color theory.
In the future, we intend to explore how we might flatten this curve through modification to the language (\eg{} via macros or simplifying concepts), structure editing~\cite{mcnutt2023projectional}, or automatic lint generation (such as via programming-by-demonstration or LLMs)---in addition to traditional learning resources.
\section{Discussion}
\label{sec:disco}

We presented a design probe that explores the complexities of visualization linters by taking color palette design as a microcosm of visualization.
We found that linters can be straightforwardly integrated into GUI-based contexts and tools, but adjustment to the affordances of those ecosystems is critical for appropriate use.
For instance, having linters ``always on'' (or visually present) is acceptable in the vast visual space of a code base, but can be overwhelming in the small plane of a color picker.
We identified a series of basic features that linters tend to have, which we used to analyze and structure our probe.

While our probe seems to be useful (but improvable), the primary contribution of this work is not this system.
Instead, we seek to answer the trio of criticisms of prior linters that opened this paper.
We first observed that linters are \textbf{primarily text-based}, which may leave out many users.
Following conventional color palette design tools, our probe was GUI-based and allowed us to demonstrate that a linter could be meaningfully integrated into a GUI tool.
While there is still ample design territory to explore, our probe suggests that this integration can be fruitful, and, more importantly, that linters need not be bound exclusively to text.
Moreover, all our participants were familiar with lint-like checks in end-user tools (\eg{} sufficient contrast in a tool like WebAIM~\cite{webAIM}) suggesting that users can be receptive to this type of integration and guidance.
While many issues are addressable via guard rails or smart defaults, complex domains like color or visualization have ample space for the subtle errors that linters excel at catching.

Next, we noted that most linters are \textbf{impolite/annoying} by presenting huge walls of errors.
Our probe explored incorporating task context (such as the palette type) and allowing users to select when and how they want linting advice---such as by focusing on examples rather than lints. We found these to be helpful ways to reduce visual cognitive load.
These issues can thus be meaningfully reduced via design rather than being intrinsically bound to the medium.
We found that opt-in control was sufficient to demonstrate a reduction in visual load, but models like probabilistic interrupts~\cite{horvitz1999principles} might reduce it further still.

Finally, we observe that linters often feature  \textbf{un-useful analysis}.
By working in a domain with richer theoretical foundations and a cleaner set of results, we found that it was possible to straightforwardly present a meaningful, complex analysis.
For instance, checking for properties like \lintRule{Even Distribution in Lightness} (\figref{fig:lint-explainer}D) as well as easy-to-miss accessibility checks like \lintRule{WCAG Contrast} (\figref{fig:lint-explainer}B).
Crucially, being able to evaluate closer to usage (than to specification) allowed us to shift our analysis focus from the interface to the interaction~\cite{beaudouin2004designing}.
To put it more simply: linters wedded to a particular domain---such as colors or choropleths or charts generally---offer a unique opportunity to validate semantics rather than merely the means they are formed in---going from \emph{is the way in which I have produced something correct?} to \emph{is what I have done correct?}
DSL tooling (\eg{} Vega-Lite's JSON Schema) typically captures most typos and surface-level errors. Instead, linters have the opportunity to surface domain knowledge by acting as proxies for the more difficult problem of semantic evaluation.

\parahead{Limitations} Like any work, ours suffers a number of limitations.
We did not experimentally explore the effect of linting on color palette design.
The purpose of our probe was to traverse the linting design space,
rather than prove that it improves palette design.
In future work we intend to compare the quality of palettes produced with related approaches.
While our interview study provides evidence that our model of color palette usage aligns with that of professional palette designers, this is not the same as an empirical evaluation using our tool.
Further, that study may be limited by selection bias as we drew from our professional networks to identify participants.
Participants spoke about their practices being common, but our sample may bias this perspective.
We attempted to circumvent these issues by using an external theory as a means of reflection (TDoPS~\cite{jakubovic2023technical}), but our results may be biased by this research-as-instrument approach.
Similarly, while our lints covered all of the topics (excluding those not supported by current theory) that participants expressed as elements that they considered in their work our survey of these lints was not conducted formally. Conducting such a survey is valuable future work.
Our probe had non-trivial conceptual and visual complexity, which may be difficult to approach for some users.
Reducing this complexity may cause some of our results to differ, but our probing of the GUI-based linter design space did not require a perfected UX.
However, we intend to continue to improve \languageName{}'s technical features and make it more approachable to non-expert users, as well as evaluate how such users (as well as designers) interact with it.
We will also explore integration of \languageName{} with systems like Figma.

\parahead{The next N linters}
This work does not try to be simply yet another linter, but instead seeks to chart a trajectory for what future works and systems in this area might look like.
Here we sketch a conceptual recipe we followed in this work, also observed in other works.

We first identified a \textbf{(1) problem domain} from which we collected appropriate guidelines, heuristics, and best practices.
It can be important during this phase to engage with a domain expert. For instance, Lei \etal{}~\cite{lei2023geolinter} engaged with a geographer (one of their paper authors).
Similarly, one of the authors of this work is a noted expert in the field of color.
We suggest that engaging with a domain expert is a valuable way to provide a layer of ecological and domain validity. Such collaboration can help answer questions like: \emph{what checks are critical?} and \emph{when is a set of checks sufficient?}
Of course, no one expert can characterize appropriate usage for an entire domain of knowledge, so validating those results (as we did in \secref{sec:formative-study}) is important.

We then selected a \textbf{(2) problem space representation} consisting of standardized color representations, such as CIELAB and RGB hex colors, enhanced with a small amount of structure (\figref{fig:pal-model}).
Similarly, tools like GeoLinter~\cite{lei2023geolinter} or VizLinter~\cite{chen2021vizlinter} use Vega-Lite~\cite{satyanarayan2016vega} as their problem space model.
Much of a tool's utility is embedded into this representation, so appropriate selection is important.
The design of these representations is critical and may benefit from drawing on theoretical models (as Vega-Lite does with the Grammar of Graphics) or expanding them (such as via a model of visualization semantics) as well as human-centered design methods (such as PLIERS~\cite{coblenz2021pliers}).

Next, we developed an \textbf{(3) assertion representation} for the problem space, namely \languageName{}---creating an instrumental interaction-style relationship~\cite{beaudouin2000instrumental} between lint and color palette.
In prior systems this representation is implicit, typically large collections of heuristics implemented in a general-purpose language. VizLinter~\cite{chen2021vizlinter} is an exception, drawing on answer-set programming.
The effectiveness of the problem and assertion representations guides the system's capability to climb the lint ladder (\secref{sec:lint-hierachy}) in a user-understandable way.
In contrast, vislint\_mpl~\cite{mcnutt2018linting} analyzed matplotlib programs, which have an inconsistent API, limiting the scope of possible checks.
Future work might further explore co-designing assertion and domain representations.

We developed a \textbf{(4) usage context} for the linter.
We found that it was useful to integrate closely with the rest of the expected usage context---which in this case was the exploratory graphical setting of \systemName{}.
Software linters are dynamically integrated throughout the code authoring experience; they are not a standalone tool used without reference or interaction with other parts of the process.
Unlike previous visualization linters, we intentionally connect ours with other forms of automation and support.
This integration offered us fertile ground on which to build better domain support while giving a broader fixing space (including both visual hints and auto-fixers).

Using this recipe and our ladder of support we suggest that linters in other domains might flourish both within visualization (covering more domains and chart types) as well as outside of it (such as for notebooks~\cite{mcnutt2023design} or \LaTeX{} documents).
For instance, being able to evaluate dashboards would be useful, such as by developing an expanded version of Qu \etals{}~\cite{qu2017keeping} work.
Similarly, an LLM prompt linter (as Henley argues for~\cite{henley24copilot}) and as a sibling to Farsight~\cite{wang2024farsight}) could be powered by Shankar \etals{} Spade~\cite{shankar2024spade}.
McNutt \etal{}~\cite{mcnutt2023design} argue for the potential of data science and notebook linters.

Across these and other domains, we suggest that the explainable and modular advice linters provide is a valuable way to evaluate the semantics of those contexts---particularly when mixed with GUIs.

\acknowledgments{%
    We thank our participants for their insights and perspectives.
    We are appreciative of reviewer's commentary, as well as the help given by Ravi Chugh, Junran Yang, Leilani Battle, and the rest of the UW IDL.
    This work was supported by the Moore Foundation.
}

\section{Supplementary Material}
\label{sec:supp}

Our materials are available at \osf{}, and include our code, a video figure, and an appendix.
Docs for our language can be found at \asLink{https://color-buddy-docs.netlify.app/}{color-buddy-docs.netlify.app}.
A live demo is at \systemURL{}

\bibliographystyle{abbrv-doi-hyperref}

\bibliography{template}

\vfill
\clearpage

\appendix{}
\section{Appendix}

\begin{figure}[t]
    \centering
    \includegraphics[width=\linewidth]{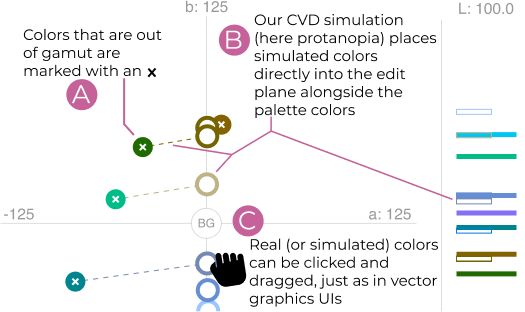}
    \vspace{-1em}
    \caption{
        \systemName{}'s edit plane supports  direct manipulation style interactions with colors---\ala{} vector manipulation tools like Figma---as well as in-situ visual explanations.
    }
    \label{fig:edit-plane}
\end{figure}

In this appendix, we provide some additional material that falls outside of the primary bounds of the paper.
In \figref{fig:edit-plane} we detail several interactions used in \systemName{} that are similar to a vector graphics manipulation tool.
In \figref{fig:color-diff} we highlight several features of \systemName{} that we did not have space to discuss in the body of this work.
Next,
we review each of the automated fixers available in our system.
Finally,
we provide a narrative use case describing how our system can support the construction of specific use palettes.

\begin{figure}[t]
    \centering
    \includegraphics[width=\linewidth]{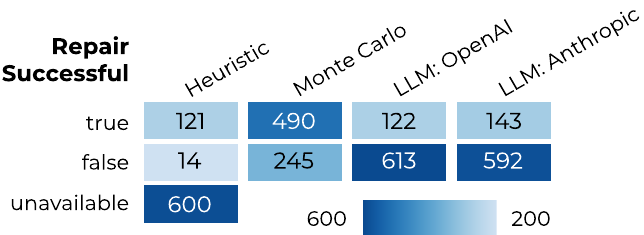}
    \caption{We ran a brief computational experiment to see which fixer performed best by running our fixers over all the length 5 Color Brewer color palettes for black and white backgrounds across 27 lints. Numbers count a failing lint fix attempt. Our Monte Carlo approach seems to perform best among these, although, subjectively, the quality of the results it produces are sometimes conceptually further away than the incorrect resolutions produced by the LLMs.}
    \label{fig:fixer-experiment}
\end{figure}

\subsection{Automated Fixers}

As described in the main text we offer several different fixers as part of our tool. These include a LLM based fixer, a set of heuristic fixers, and a Monte Carlo-inspired search algorithm.
These fixers are available to the user once a lint has failed. The user can press the ``fixes'' button next to the lint failure (as in \figref{fig:fix-menu}) where they have options to generate fixes using several different techniques.
Here we detail each of those techniques and describe an experiment we did to compare them.

We stress that all three of these techniques are naive and we do not offer them individually as contributions.
Instead, we list them here for completion.
More sophisticated methods would likely perform better.
For instance, a more robust sampling technique, such as rejection sampling, or a global optimization technique, such as simulated annealing, would likely yield higher-quality results. However, the contribution of this work centers around demonstrating that these types of interactions are possible (and integrable with a GUI), rather than identifying the best strategies for components of the work.
In future work we intend to investigate better-optimized algorithms (such as via more elegant sampling techniques) or by applying ideas such as from program repair to identify and adjust what parts of the program are failing.

\subsubsection{LLM-based Fixer}

We implemented  a simple method for fixing lints via LLMs.
We do so by formatting the string summarizing the error (which each lint produces) and that palette as a prompt to an LLM. We developed our system Winter and Spring of 2024, and used OpenAI GPT4o and Anthropic Haiku. Here is an example of a prompt:

\begin{quote}
    You are a color expert. You take in a color palette and an error it has and fix it. Your fixes should be clever but respect the original vibe of the palette. Present your fixes as a single JSON object that describes the color palette. It should have a type like {"background": string; "colors": string[]}. \\
    Additional criteria:\\
    - As much as possible, do not provide fixes by simply removing a color from the palette. \\
    - DO NOT JUST RETURN THE SAME COLORS. That is not a fix. You must change at least one color.

    Do not offer any other response. YOU MUST GIVE A VALID JSON OBJECT. If you do not, you will be banned.

    Palette: {\hlc[green!30]{["\#0084ae","\#e25c36","\#8db3c7","\#e5e3e0","\#eca288"]}}\\
    Context: {\hlc[yellow!30]{"This is a diverging palette called 'new palette'."}}\\
    Background Color: {\hlc[green!30]{"\#fff"}}\\
    Error: {\hlc[red!30]{"These colors (\#8db3c7, \#e5e3e0, \#eca288) do not have a sufficient contrast do not have sufficient contrast with the background to be easily readable.}}

    Failed: {\hlc[blue!30]{ALL a IN colors SUCH THAT contrast(a, background, WCAG21) > 3"}}
\end{quote}

This prompt is a simple string template parameterized by the {\hlc[green!30]{palette}}, a summary of the {\hlc[yellow!30]{palette's metadata}}, a {\hlc[red!30]{text summary of the error}}, and the {\hlc[blue!30]{SQL-style representation of the lint}}.
GPT4o responded to this particular prompt with a suggestion to change the palette to be \#0084ae, \#e25c36, \#5a7d91, \#b3b0ad, \#d17a4a with background \#fff.

\subsubsection{Heuristic Fixes}
\label{sec:heuristics}

We implemented a collection of heuristic fixes for a subset (7) of the built in lints. These operate outside of \languageName{} and are not parameterized by a given lints configuration. Here we give the conceptual operations for each fix---for more detail, see the code repository. These are naive heuristics meant only to solve these issues in common cases, rather than be definitive fixes.
To wit, the Sequential Palette fix uses an incorrect approximation to the full complication of perceptual ordering~\cite{bujack2018ordering}. More robust solutions are available (\cf{} Nardini \etal{}~\cite{nardini2021automatic}), but these are meant to be lightweight proofs-of-concept.

\paragraph{Diverging Palette In Appropriate Order (Definitional)} Generate all possible orders of colors, select one that is diverging.

\paragraph{Color Name Discriminability~\cite{heer2012color}} Measure the distance from each color to average color centers for a list of color names in LAB. Remove those not in conflict, but maintain a list of in-use color names. In palette order assign names that closest match for each color, maintaining a list of assignments. Use assignments to filter possible matches. Continue until all colors have assigned names. For those that changed names, use the color center as the new color value.

\paragraph{Even Distribution in Hue~\cite{bujack18GoodBadUgly} } Distribute the points evenly in hue from LCH using piece-wise linear interpolation, keeping the endpoints fixed.

\paragraph{Even Distribution in Lightness~\cite{bujack18GoodBadUgly} } Distribute the points evenly in lightness from LCH using piece-wise linear interpolation, keeping the endpoints fixed.

\paragraph{In Gamut~\cite{bujack18GoodBadUgly}} Convert the colors to hex representation and then back to the original color space.

\paragraph{Max Colors~\cite{Muth22Color}} Remove colors from the end of the palette until the maximum is reached.

\paragraph{Sequential Palette In Appropriate Order (Definitional)} Sort the colors by L* in LAB.

\subsubsection{Monte Carlo-inspired Fixer}

Finally, we implemented a Monte Carlo-inspired fixer as a way to integrate the specific configuration of \languageName{} programs into the fixing process. For a given failing lint, we identify which colors in a palette are to blame (using our standard blame technique), call them $colors_{blamed}$. For each color in  $colors_{blamed}$ we randomly perturb that color in whatever space the color is specified. The perturbation consists of a step in each of the color space's axes of random length with a unit maximum step. For instance, in LAB a color might receive a +0.8 L, -0.3a*, +0.1b*. After perturbation, the updated palette is checked against the original lint. If it is passing then the process stops, if not then the process repeats. This continues until either a solution is found, or a pre-specified number of rounds is hit.

\subsubsection{Fix Repair Effectiveness}

While these fixers are meant to only demonstrate that fixers are possible in this context, we also sought to get a typical sense of whether or not the results produced  by the fixers did so successfully, and, moreover, which of them was the most successful.
To do so we conducted a small computational experiment to gauge how well their repairs worked for a typical workload. To simulate this we formed a dataset consisting of all Color Brewer color palettes of length 5 on both black and white backgrounds. We selected this number because this is the number of colors by which Color Brewer represents each of the palettes, suggesting that it is a typical volume---although this is a limitation of this experiment.
Several built-in lints (\figref{fig:lint-list}) pass trivially on this dataset. These include all of the affect rules, rules relating to text, and so on. This yields 35 - 8 = 27 relevant lints.

We summarize the results of this experiment in \figref{fig:fixer-experiment}.
We found that the Monte Carlo strategy performed best in general, however this should be unsurprising because this is the only approach that involved a direct connection to the particular lint under study.
We emphasize that this is an informal experiment meant only to provide a back-of-the-envelope evaluation of this feature.
The details of this experiment are also presented in our supplemented materials.

\subsection{Case Study: Tweaking Brand Colors}
Katy is a data engineer at a large city's transit organization. As part of her duties, she often has to prepare dashboards describing analytics, such as trips delivered on time, the state of the fleet, and so on. As with many transit systems, theirs has a set of color-coded lines that colors serve as the name of the route (such as the blue or green line). Because of this semantic relationship, charts in the dashboards she makes often need to present data about particular routes using the corresponding colors for those routes---for instance using the blue-line blue from the organization's style guide.
Unfortunately, these colors are poorly suited to visualization, as they are primarily focused on signage discriminability.
Katy wants to modify this set of colors in a way that is better for visualization but still in line with the brand.

She uploads the colors to \systemName{} and navigates to the evaluation page. She sees that things are generally okay, and so flips to the examples section.
Her organization exclusively uses line charts and bar charts to accommodate the variety of literacy levels in her organization. The one exception is string-line charts (also called Marey train schedule diagrams), appropriate for the agency's focus on trains.
\systemName{} does not have this chart type built-in (\figref{fig:lint-list}), so she uploads an existing SVG string-line to the examples page.

\begin{figure}[t]
    \centering
    \includegraphics[width=\linewidth]{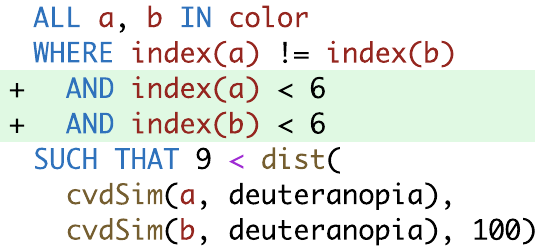}
    \caption{Modifying a lint to check only the first 6 colors in a palette.}
    \label{fig:prog-diff}
\end{figure}

Watching the examples closely, she makes a series of adjustments, making the yellow less bright so that it stands out less on the string-line, darkening the blue so it is less vibrant in comparison with the other bars in a bar chart, and a host of other small tweaks.
She then flips back to the evaluation page and notices that there are now a handful of errors---the new colors are not CVD-friendly for Deuteranopia and Tritanopia, inappropriate contrast with the background, and it is hard to tell the difference between the gray and the brown for thin lines.
She knows that getting the accessibility checks right is especially important for work in the government.
It is okay for her palette to have some dynamic range, so she turns off the \lintRule{Fair} warning by clicking Ignore.
Her new yellow is blamed for the inappropriate contrast, so she increases the lightness until the check passes.
After trying out a couple of possible fixes from the AI and making some manual adjustments, she is having a hard time getting all 10 of the colors to work for Deuteranopia.
She observes that most of the time only the first few colors are going to be used, so she decides that, for this palette only the first six colors need to be CVD-friendly (\pxx{1} and \pxx{2} described using this technique in their palette designs).
For each CVD type, she adds an AND predicate to the where clause, as in \figref{fig:prog-diff}, applying the lint to only the first 6 colors.

\begin{figure*}[t]
    \centering
    \includegraphics[width=\linewidth]{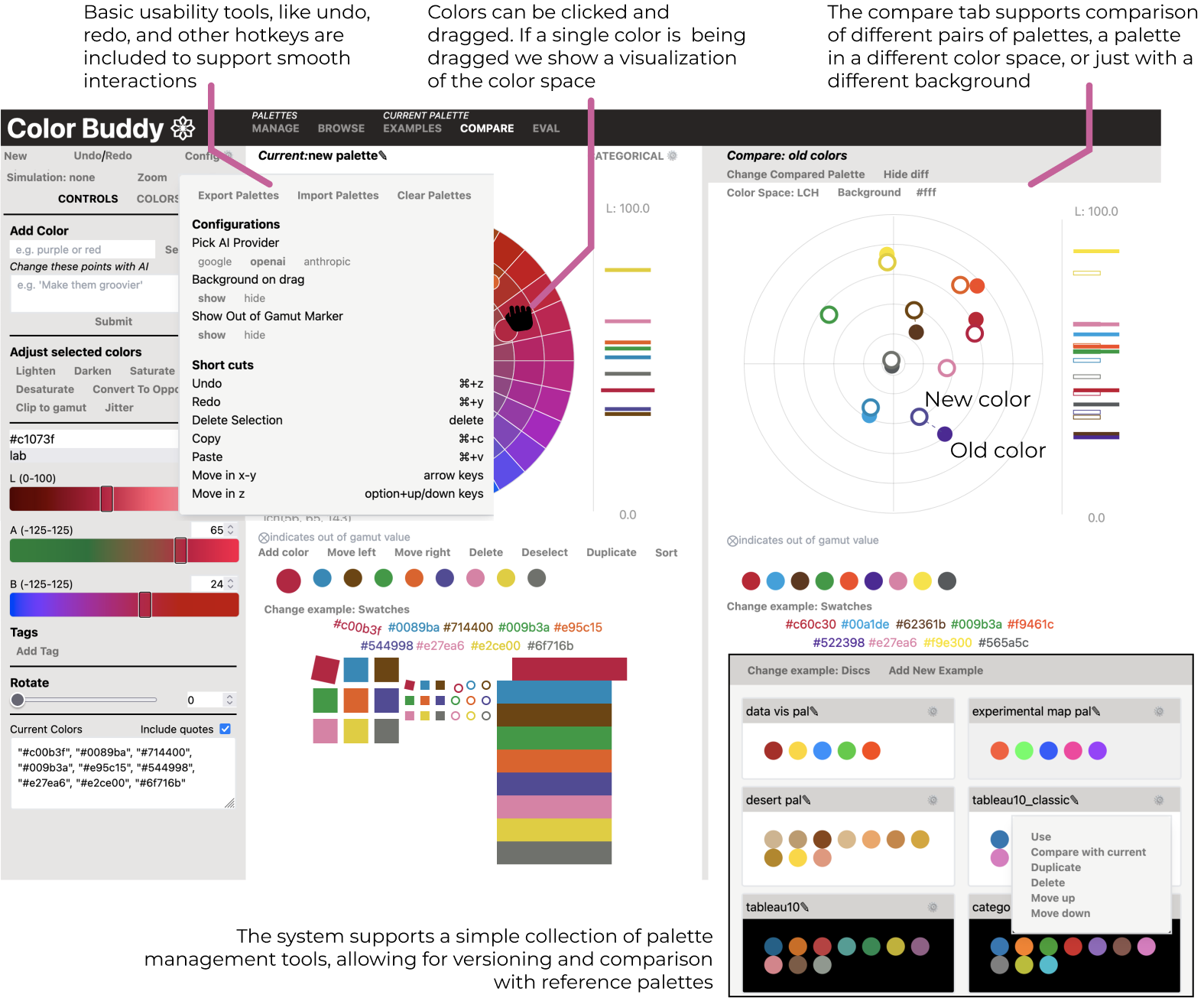}
    \caption{\systemName{} includes additional features not described in the main paper. These include functionality highlighted in the case study, such as the ability to visualize the difference between two palettes using the same visual metaphor as our CVD simulator. In addition it also includes basic usability tools for palette design, such as state management (\ie{} undo and redo) and  palette management (supporting versioning).}
    \label{fig:color-diff}
\end{figure*}

She wants to make sure that the colors did not change too much, so she creates another palette with the original brand colors.
She then goes to the Compare section and diffs the two versions (\figref{fig:color-diff}). Most colors have not significantly changed---everything is still recognizable as its original color---but everything is more legible.
There are a few more issues in the Eval section but she is happy with what she has so she turns off the rest of the warnings and exports the palette.

\end{document}